\documentclass[twocolumn,preprintnumbers,amsmath,amssymb,showpacs]{revtex4}
\usepackage{dcolumn}
\usepackage{bm,epic,eepic}
\usepackage[english]{babel}
\usepackage{amsfonts}
\usepackage{amsthm}
\usepackage{dsfont}
\usepackage{graphicx}
\usepackage{pst-plot}
\usepackage{pstricks}
\usepackage{pst-coil}

\newcommand{\av}[1]{\langle #1 \rangle}
\newcommand{\BEQ}{\begin{eqnarray}}
\newcommand{\EEQ}{\end{eqnarray}}
\newcommand{\ket}[1]{ | \, #1 \rangle}

\newcommand{\bra}[1]{ \langle \, #1  |}
\renewcommand{\H}{\mathcal{H}}
\newcommand{\beq}{\begin{equation}}
\newcommand{\eeq}{\end{equation}}
\newcommand{\Eq}[1]{Eq.\,\,(\ref{#1})}

\newcommand{\forget}[1]{}

\renewcommand{\vec}{\bm}

\begin{document}

\title{Classification and monogamy of three-qubit biseparable Bell correlations
}
\author{Michael Seevinck}
\email{seevinck@phys.uu.nl}
\affiliation{%
Institute for History and Foundations of Science,\\
 Utrecht University
 PO Box 80.000, 3508 TA Utrecht, the Netherlands}%

\date{\today}

\begin{abstract}
We strengthen the set of Bell-type inequalities presented by Sun \& Fei
[Phys. Rev. A {\bf 74}, 032335 (2006)]
that give a classification for biseparable correlations and entanglement in tripartite quantum systems.
We will furthermore consider the restriction to local orthogonal spin observables
and show that this strengthens all previously known such tripartite inequalities.
The quadratic inequalities we find indicate a type of monogamy of maximal biseparable tripartite  quantum correlations, although the nonmaximal ones can be shared.
This is contrasted to recently found monogamy inequalities for bipartite Bell correlations in tripartite systems.

\end{abstract}
\maketitle
\section{Introduction}
\noindent
Although Bell inequalities have been originally proposed to test quantum
mechanics against local realism, nowadays  they also serve another purpose, namely investigating
quantum entanglement. Indeed, Bell inequalities were  used to give detailed characterisations
of multipartite entangled states by giving bounds on the correlations that these states can give
rise to \cite{a}.

Recently a set of Bell-type inequalities was presented by Sun \& Fei \cite{sun} that gives a finer
classification for entanglement in tripartite systems than was previously known. 
The inequalities
distinghuish three different types of bipartite entanglement that may exist in tripartite systems. 
They not only determine if one of the three parties is separable with respect to the
other two, but also which one. It was shown that the three inequalities give a bound that 
can be thought of as
tracing out a sphere in the space of expectations of the three Bell operators that were used in
the inequalities.  Here we strenghten this bound 
by showing that all states are confined within the interior of the intersection of three cylinders and
the already mentioned sphere.

Furthermore, in Refs. \cite{roy,uffink06} it was shown that considerably stronger separability
inequalities for the expectation of  Bell operators can be obtained if one restricts oneselves
 to local orthogonal spin observables (so-called LOO's\cite{loos}).
We will show that the same is the case for the Bell operators considered here by strengthening
all above mentioned tripartite inequalities under the restriction of using
orthogonal observables.

The relevant tripartite inequalities are included in the $N$-particle inequalities derived in
\cite{chen}. It was shown that these $N$-partite inequalities can be violated maximally by the
$N$-particle maximally entangled Greenberger-Horne-Zeilinger (GHZ) states\cite{chen}, but, as will be shown here, they can
also be maximally violated by states that contain only $(N-1)$-partite entanglement.  Although
these inequalities thus allow for further classification of multipartite entanglement (besides
some other interesting properties), they can not be used to distinghuish
full $N$-particle entanglement from $(N-1)$-particle entanglement in $N$-partite states. It is
shown that this is neither the case for the stronger bounds that are derived for the case of
LOO's.

In section II the analysis for unrestricted spin observables is performed and in 
section III for the restriction to LOO's. Lastly, in the discussion of section IV we will interpret the presented quadratic inequalities 
as indicating a type of monogamy 
of maximal biseparable three-particle quantum correlations. 
Nonmaximal correlations can however be shared. This is contrasted to the recently
found monogamy inequalities of Toner \& Verstraete \cite{toner}.

\section{Unrestricted observables}\noindent 
Chen et al. \cite{chen} consider $N$-parties that each have two alternative dichotomic measurements
denoted by $A_j$ and $A_j'$  (outcomes $\pm1$) and show that local realism (LR) requires that
\beq\label{ineqN}
|\av{D_{\mathrm{LR}}^{(i)}}|= \frac{1}{2}|\av{B_{N-1}^{(i)}(A_i +A_i ') +(A_i -A_i ')}_{\mathrm{LR}} |\leq1,
\eeq
for $i=1,2,\ldots, N$, where $B_{N-1}^{(i)}$  is the Bell polynomial of the Werner-Wolf-\.Zukowski-Brukner (WWZB)
inequalities\cite{werner} for the $N-1$ parties, except for party $i$.
These Bell-type inequalities have only two different local settings and  are contained in the general inequalities for $N>2$ parties that have more than two alternative measurement settings derived by Laskowski et. al \cite{laskowski}.
Indeed, they follow from the latter when choosing certain settings equal.  Note furthermore that the WWZB
inequalities are contained in the inequalities of \Eq{ineqN} by choosing $A_N=A_N'$.

The quantum mechanical counterpart of the Bell-type inequality of \Eq{ineqN}  
is obtained by introducing
dichotomic observables $A_k$, $A_k'$ for each party $k$. Let us define analogously  to Sun \& Fei \cite{sun} the operator
\beq\label{qmcounter}
\mathcal{D}_N ^{(i)}=\mathcal{B}_{N-1}^{(i)}\otimes(A_i +A_i')/2 + \openone_{N-1}^{(i)}\otimes(A_i -A_i')/2,
\eeq
for $i=1,2,\ldots, N$. Here $\mathcal{B}_{N-1}^{(i)}$
and $ \openone_{N-1}^{(i)}$ are respectively
the Bell operator of the WWZB inequalities and the identity operator both for the $N-1$ qubits not involving qubit $i$.

The quantum mechanical counterpart of the local realism inequalities of
\Eq{ineqN}  for all $i$ is then
\beq\label{ineqNqm}
|\av{\mathcal{D}_N ^{(i)}}|\leq 1,
\eeq
where $\av{\mathcal{D}_N ^{(i)}}:=\mathrm{Tr}[\mathcal{D}_N ^{(i)}\rho]$ and
$\rho$ is a $N$-party quantum state.

Since the Bell inequality of \Eq{ineqNqm} uses only two alternative dichotomic observables for each party the maximum violation of this Bell inequality is obtained for an $N$-particle pure qubit state and furthermore for projective observables, as proven recently by Masanes \cite{masanes} and by Toner \& Verstraete\cite{toner}. 
In the following we will thus consider qubits only and the observables will be represented by the spin operators
$A_k=\vec{a}_k\cdot \vec{\sigma}$ and $A_k'=\vec{a}'_k\cdot \vec{\sigma}$ with $\vec{a}_k$ and $\vec{a}_k '$
unit vectors that denote the measurement settings and
$\vec{a}\cdot \vec{\sigma}=\sum_l a_l\sigma_l$ where $\sigma_l$ are the familiar Pauli spin observables for $l=x,y,z$ on $\H=\mathbb{C}^2$. In fact, it suffices \cite{toner} to consider only real and traceless observables, so we can set $a_y=0$ for all observables.

An interesting feature of the inequalities in \Eq{ineqNqm} is that all
generalised GHZ states $\ket{\psi_\alpha^N}=\cos \alpha \ket{0}^{\otimes N} +\sin
\alpha \ket{1}^{\otimes N}$ can be made to violate them 
for all $\alpha$ \cite{chen,laskowski}, which is not the case for the WWZB inequalities. Furthermore,
the maximum is given by
\beq\label{maxn}
\max_{A_i,A_i'}|\av{\mathcal{D}_N ^{(i)}}|=
 2^{(N-2)/2},\eeq  as was
proven by Chen et al. \cite{chen}. They also noted that this maximum is
obtained for the maximally entangled $N$-particle GHZ state $\ket{GHZ_N}$
(i.e., $\alpha=\pi/4$) and for all local unitary transformations of this state.
However, not noted in \cite{chen} is the fact that the maximum is also
obtainable by $N$-partite states that only have $(N-1)$-particle entanglement,
which is the content of the following theorem.

\emph{Theorem 1.} Not only can the maximum value of $2^{(N-2)/2}$ for
$\av{\mathcal{D}_{N}^{(i)}}$ be reached by fully $N$-particle entangled states
(proven in \cite{chen}) but also by $N$-partite states that only have
$(N-1)$-particle entanglement.

\emph{Proof}: Firstly, $(\mathcal{B}_{N-1}^{(i)})^2\leq2^{(N-2)} \openone_{N-1}^{(i)}$ (as
proven in \cite {werner}). Here $X\leq Y$ means that $Y-X$ is semipositive definite. Thus the
maximum possible eigenvalue of $\mathcal{B}_{N-1}^{(i)}$
is $2^{(N-2)/2}$. Consider a state $|
\Psi_{N-1}^{(i)}\rangle$ for which
$\av{\mathcal{B}_{N-1}^{(i)}}_{\ket{\Psi_{N-1}^{(i)}}}=2^{(N-2)/2}$. This must
be \cite{werner} a maximally entangled $(N-1)$-particle state (for the $N$
particles except for particle $i$), such as the state $\ket{GHZ_{N-1}}$.
Next consider the state
$\ket{\xi^{(i)}}=\ket{\Psi_{N-1}^{(i)}}\otimes\ket{0_i}$, with $\ket{0_i}$ an
eigenstate of the observable $A_i$ with eigenvalue $1$. This is an $N$-partite
state that only has $(N-1)$-particle entanglement. Furthermore choose
$A_i=A'_i$ in \Eq{qmcounter}. We then obtain
$\av{\mathcal{D}_{N}^{(i)}}_{\ket{\xi^{(i)}}}=
\av{\mathcal{B}_{N-1}^{(i)}}_{\ket{\Psi_{N-1}^{(i)}}}\av{A_i}_{\ket{0_{i}}}=2^{(N-2)/2}$,
which was to be proven. ${\scriptsize{\square}}$

This theorem thus shows that the Bell inequalities of \Eq{ineqNqm} can not
distinghuish between full $N$-partite entanglement  and $(N-1)$-partite
entanglement, and thus can not serve as full $N$-particle entanglement
witnesses.

Let us now concentrate on the tri-partite case ($N=3$ and $i=1,2,3$). Sun \&
Fei \cite{sun} obtain that for fully separable three particle states 
it follows that $|\av{\mathcal{D}_3
^{(i)}}|\leq 1$, which does not violate the local
realistic bound of \Eq{ineqN}.  General three particle states give
$|\av{\mathcal{D}_3^{(i)}}| \leq \sqrt{2}$, which follows from \Eq{maxn}. As
follows from Theorem 1  this can be saturated
by both fully entangled three particle states as well as for bi-separable
entangled three particle states (e.g., two-partite entangled three particle
states).

Sun \& Fei have furthermore presented a set of Bell inequalities that
distinghuish three possible forms of bi-separable entanglement. They consider
biseparable states that allow for the partitions $1-23$, $2-13$ and $3-12$
where the set of states in these partitions is denoted as $S_{1-23}, S_{2-13},
S_{3-12}$ and which we label by $j=1,2,3$ respectively. These sets contain
states such as $\rho_{1}\otimes\rho_{23},~\rho_{2}\otimes\rho_{13}$, and
$\rho_{3}\otimes\rho_{12}$ respectively.
For states in partition $j$  (and for $i=1,2,3$) Sun \& Fei obtained
 \begin{align}
|\av{\mathcal{D}_3^{(i)}}|\leq \chi_{i,j}\,,
 \end{align}
 with $\chi_{i,j}=\sqrt{2}$ for $i=j$  and $\chi_{i,j}=1$ otherwise.

 They furthermore proved that for all three qubit states
   \beq\label{4}
 \av{\mathcal{D}_3^{(1)}} ^2+\av{\mathcal{D}_3^{(2)}}^2+
 \av{\mathcal{D}_3^{(3)}}^2 \leq 3,~~~\forall \rho.
 \eeq
   Although this inequality is stronger than the set above (for details see Fig. 1 in \cite{sun}),
   it can be saturated by fully separable states. For example, choose the state $\ket{000}$ and
   choose all observables to be projections 
   onto this state. Then we get
   $\av{\mathcal{D}_3^{(1)}}_{\ket{000}} ^2+\av{\mathcal{D}_3^{(2)}}_{\ket{000}}^2+
   \av{\mathcal{D}_3^{(3)}}_{\ket{000}}^2 =3$.

\begin{figure}[!h]
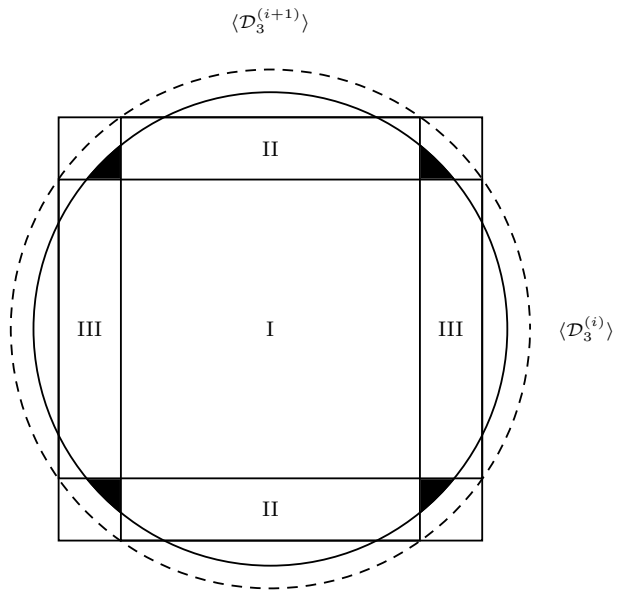

\pspicture(-3,-3.7)(3,4.6)
\psset{xunit=2,yunit=2}
\pswedge[fillstyle=solid,fillcolor=black](0,0){3.162}{38}{52}
\pswedge[fillstyle=solid,fillcolor=black](0,0){3.162}{128}{142}
\pswedge[fillstyle=solid,fillcolor=black](0,0){3.162}{218}{232}
\pswedge[fillstyle=solid,fillcolor=black](0,0){3.162}{308}{322}
\psframe*[linewidth=0.7pt,linecolor=white,fillcolor=white](-1,-1.414)(1,1.414)
\psframe*[linewidth=0.7pt,linecolor=white,fillcolor=white](-1.414,-1)(1.414,1)
\psframe[linewidth=0.7pt,fillcolor=white](-1,-1.414)(1,1.414)
\psframe[linewidth=0.7pt,fillcolor=white](-1.414,-1)(1.414,1)
\psframe[linewidth=0.7pt](-1.414,-1.414)(1.414,1.414)
\pscircle[linestyle=dashed,dash=4pt 3pt,linewidth=0.7pt](0,0){3.464}
\pscircle[linewidth=0.7pt](0,0){3.162}
\rput(0,0){\footnotesize{I}}
\rput(0,1.2){\footnotesize{II}}
\rput(0,-1.2){\footnotesize{II}}
\rput(1.2,0){\footnotesize{III}}
\rput(-1.2,0){\footnotesize{III}}
\rput(0.0,2.05){\scriptsize{$\av{\mathcal{D}_3^{(i+1)}}$}}
\rput(2.1,0){\scriptsize{$\av{\mathcal{D}_3^{(i)}}$}}
\endpspicture
\caption{$\mathcal{D}_3^{(i)}-\mathcal{D}_3^{(i+1)}$ plane with the stronger bound given by
the circle with radius $\sqrt{5/2}$ which strengthens the less strong bound with radius $\sqrt{3}$ that is
given by the dashed circle.}
\label{fig1}
\end{figure}

   Let us consider $\mathcal{D}_3^{(i)}$ (for $i=1,2,3$) to be three coordinates of a space
   in the same spirit as Sun \& Fei \cite{sun} did.
They showed that the  fully separable states are confined to a cube with edge length $2$ and the
biseparable states in partition $j=1,2,3$ are confined to cuboids with size
either$ 2\sqrt{2}\times2\times2 ,~ 2\times  2\sqrt{2}\times2$, or
$2\times2\times2\sqrt{2}$.  Note that states exist that are biseparable with respect to all three partitions (and thus must lie within the cube with edge length $2$), but which are not fully separable \cite{bennett}. Furthermore, all three-qubit states are in the
intersection of the cube with size $2\sqrt{2}$ and of the sphere with radius
$\sqrt{3}$. Sun \& Fei note that this sphere is just the external sphere of the
cube with edge $2$, which is consistent with the above observation that fully
separable states can lie on this sphere. If we look at the
$\mathcal{D}_3^{(i)}-\mathcal{D}_3^{(i+1)}$ plane we get Fig. \ref{fig1}. The
fully separable states are in region I; region II belongs to  the biseparable
states of partition $j=i+1$; and region III belongs to states of partition
$j=i$. Other biseparable states and fully entangled states are outside these
regions but within the circle with radius $\sqrt{3}$. 
   However, in the following theorem we show a quadratic inequality even stronger than
   \Eq{4} which thus strengthens the bound in Fig. \ref{fig1} given by the circle of radius $\sqrt{3}$
and which forces the states just mentioned into the black regions.

\emph{Theorem 2}.
  For the case where each observer chooses between two settings all three qubit states
  obey the following inequality:
  \beq\label{new}
  \av{\mathcal{D}_3^{(i)}} ^2+\av{\mathcal{D}_3^{(i+1)}}^2\leq  \frac{5}{2},~~~\forall \rho,
 \eeq
 for $i=1,2,3$ and where $i$ and $i+1$ are both modulo 3.

 \emph{Proof}: The proof uses the exact same steps of the proof of \Eq{4} as performed by Sun \& Fei
 (i.e., proof of Theorem $2$ in \cite{sun}) and can be easily performed for the left hand side
 of \Eq{new} that contains only two terms instead of the three terms on the right hand side of
 \Eq{4}.  This results in only  a minor change in calculations \footnote{
 In further detail,
steps (1) to (4) of the proof in Sun \& Fei \cite{sun} become (using the terminology of their proof): 
 (1):  $\omega = 2(\vec{s}_1\otimes\vec{s}_2\otimes\vec{s}_3\cdot \vec{Q})^2=2\bra{\Psi}C_1C_2C_3\ket{\Psi}^2\leq 2$,  (2):  $\omega = 2(\vec{s}_1\otimes\vec{s}_2\otimes\vec{s}_3\cdot \vec{Q}+\vec{s}_1\otimes\vec{s}_2\otimes\vec{t}_3\cdot \vec{Q})^2=2\bra{\Psi}C_1C_2(C_3+D_3)\ket{\Psi}^2\leq 2$,  
 (3): $\omega=(5/4)(\cos(\theta_1+\theta_2+\theta_3) -\sin(\theta_1+\theta_2+\theta_3))^2\leq5/2$, 
 (4):  $\omega=(\cos(\theta_1+\theta_2+\theta_3) -\sin(\theta_1+\theta_2+\theta_3))^2\leq2$.
Here $\omega=\av{\mathcal{D}_3^{(i)}} ^2+\av{\mathcal{D}_3^{(i+1)}}^2$ (i.e., the l.h.s. of \Eq{new}), where  we have chosen $i =1$. Note that by symmetry the proof goes analogous for $i=2,3$. It follows that step (3) has the highest bound of $5/2$.}.  Case (3) in this proof then has the highest bound of $5/2$, whereas  the other three cases give a lower bound equal to $2$.   
 $\square$
 
Note that in contrast to \Eq{4} the inequality of \Eq{new} can not be saturated by separable
states, since the latter have a maximum of 2 for the left hand expression in \Eq{new}.

If we again look at the space  given by the coordinates $\mathcal{D}_3^{(i)}$ (for $i=1,2,3$),
we have thus found that all states are, firstly, confined within the intersection of the three
orthogonal cylinders
$\av{\mathcal{D}_3^{(i)}} ^2+\av{\mathcal{D}_3^{(i+1)}}^2\leq 5/2$ (with $i+1$
and $i+2$  both modulo $3$)
each with radius $\sqrt{5/2}$ and, secondly, they must furthermore still lie within the cube
of edge length $2\sqrt{2}$, and thirdly they must also lie within the sphere with radius $\sqrt{3}$.
In Fig. \ref{fig1} we see the strengthened bound of \Eq{new} as compared to the bound of Sun
\& Fei. 
However,  we see from this figure that
neither the intersection of the three cylinders, nor the sphere,
nor the cube give tight bounds.

The black areas in Fig. \ref{fig1} are nonempty.
For the case of \Eq{new} states thus exist that have both $|\av{D_{3}^{(i)}}|>1$ and $|\av{D_{3}^{(i+1)}}|>1$ 
(for some $i$). For example, the so-called $W$-state
\beq \label{W}
\ket{W}=(\ket{001} +\ket{010} +\ket{100})/\sqrt{3},
\eeq
 gives  $|\av{D_{3}^{(i)}}|=1.022$ for all $i$  when the observables are chosen as follows:  $A_i= \cos\alpha_i \, \sigma_z +\sin\alpha_i\,\sigma_x$  with $ \alpha_i= -0.133 $  and $A_i' = \cos\beta_i\,  \sigma_z +\sin\beta_i\,\sigma_x$
with $ \beta_i =0.460$.
 
\section{Restriction to local orthogonal spin observables}
\noindent In Refs. \cite{roy,uffink06} it was shown that considerably stronger separability
inequalities for the expectation of the bipartite Bell operator
$\mathcal{B}_2$ can be obtained if one restricts oneselves to local orthogonal observables (LOO's).
We will now show that the same is the case for the Bell operator $\mathcal{D}_3^{(i)}$. 
The following theorem strenghtens all previous bounds of section {II} for general observables.

\emph{Theorem 3}.
Suppose all local observables are orthogonal, i.e., $\vec{a}_i\cdot\vec{a}_i'=0$,
then the following inequalities hold:
\begin{description}
\item (i) For all states: $|\av{\mathcal{D}_{3}^{(i)}} |\leq \sqrt{3/2}\approx1.225$.
\item (ii) For fully separable states:  $|\av{\mathcal{D}_{3}^{(i)}} |\leq\sqrt{3/4}\approx0.866$.
\item (iii) For biseparable states in partition $j=1,2,3$:
 \begin{align}
|\av{\mathcal{D}_3^{(i)}}|\leq \chi_{i,j}\,,
 \end{align}
 with $\chi_{i,j}=\sqrt{3/2}\approx1.225$ for $i=j$  and $\chi_{i,j}=\sqrt{3/4}\approx0.866$ otherwise.
\item (iv) Lastly,
 for all states:
\beq\label{new2}
\av{D_{3}^{(i)}}^2 +\av{D_{3}^{(i+1)}}^2  \leq 2.\eeq
\end{description}

\emph{Proof}:
(i) The square of $\mathcal{D}_3 ^{(i)}$ is given by
\beq\label{square}
(\mathcal{D}_3 ^{(i)})^2=(\mathcal{B}_{2} ^{(i)})^2
\otimes\frac{1}{2}(1+\vec{a}_i\cdot\vec{a}'_i) \openone_i
+ \openone_{2}^{(i)}\otimes\frac{1}{2}(1-\vec{a}_i\cdot\vec{a}'_i) \openone_i,
\eeq
where $ \openone_{2}^{(i)}$ is the identity operator for the $2$ qubits not
including qubit $i$. For orthogonal observables we get $\vec{a}_i\cdot\vec{a}'_i=0$, and
$(\mathcal{B}_{2} ^{(i)})^2\leq 2 \openone_2^{(i)}$ (as proven in \cite{roy,uffink06}). 
The maximum eigenvalue of $(\mathcal{D}_3 ^{(i)})^2$ is thus $3/2$, which implies that
$|\av{\mathcal{D}_{3} ^{(i)}}|\leq \sqrt{3/2}$.

(ii) For fully separable states we have from \Eq{qmcounter} that
\beq\label{proofi}
\av{\mathcal{D}_3 ^{(i)}}=\frac{1}{2}(\av{\mathcal{B}_{2}^{(i)}} \av{(A_i +A_i')} +\av{(A_i -A_i')}).
\eeq
Furthermore for the case of orthogonal observables $|\av{\mathcal{B}_{2}^{(i)}}| \leq 1/\sqrt{2}$
\cite{roy, uffink06}. Thus $|\av{\mathcal{D}_3 ^{(i)}}|\leq|( \av{(A_i +A_i')}/\sqrt{2} +\av{(A_i -A_i')})/2|.$
Since the averages are linear in the state $\rho$ the maximum is obtained for a pure state of qubit $i$.
This state can be represented as $1/2( \openone +\vec{o}\cdot\vec{\sigma})$, with $|\vec{o}|=1$ and
$\vec{o}\cdot\vec{\sigma}=\sum_k o_k \sigma_k$ ($k=x,y,z$).
Take $C=(A_i +A_i')$, $D=(A_i -A_i')$ and
$\vec{s}=\vec{a}_i+\vec{a}_i'$, $\vec{t}=\vec{a}_i-\vec{a}_i'$. We get $|\vec{s}|=|\vec{t}|=\sqrt{2}$.
Choose now without losing generality \cite{toner}
 $\vec{s}=\sqrt{2}(\cos\theta,0,\sin\theta)$ and $\vec{t}=\sqrt{2}(-\sin\theta,0,\cos\theta)$.
 Then
 \begin{align}
| \av{\mathcal{D}_3 ^{(i)}}|&\leq |( \vec{s}\cdot\vec{o}/\sqrt{2} +  \vec{t}\cdot\vec{o} )/2|\nonumber\\
&= |\frac{1}{2}\big((o_z-\sqrt{2}o_x)\sin\theta +(o_x+\sqrt{2}o_z)\cos\theta|\big).\nonumber
  \end{align}
 Maximizing over $\theta$ (i.e., $\max_\theta (X\cos\theta +Y\sin\theta) =\sqrt{X^2 +Y^2}$) and
 using $o_x^2 +o_y^2+o_z^2=1$
 we finally get
  \begin{align}
| \av{\mathcal{D}_3 ^{(i)}}|&\leq |\sqrt{3/4(o_x^2 +o_z^2)}|\leq\sqrt{ 3/4}
.
  \end{align}

(iii)  For biseparable states in partition $j=i$ we get the same as in \Eq{proofi}, but now
  $|\av{\mathcal{B}_{2}^{(i)}}| \leq \sqrt{2}$. Using the method of (ii) we get
  \begin{align}
| \av{\mathcal{D}_3 ^{(i)}}|&\leq| ( \sqrt{2}\,\vec{s}\cdot\vec{o} +  \vec{t}\cdot\vec{o} )/2|
\leq\sqrt{3/2}
.
  \end{align}
For biseparable states in partition $i+1$ and $i+2$ a somewhat more elaborous proof is needed.
Let us set $i=1$ and $j=3$ for convenience (for the other partition $j=2$ we get the same result).
The maximum is again obtained for pure states. Every pure state in partition $j=3$ can be written as
$\ket{\psi}=\ket{\psi}_{12}\otimes\ket{\psi}_3$. Then
\begin{align}
|\av{\mathcal{D}_3 ^{(i)}}|=&\,|        \frac{1}{4}\av{(A_1+A_1')(A_2+
A_2')} _{\ket{\psi_{12}}}\av{A_3}_{\ket{\psi_3}}\nonumber\\&
     +\frac{1}{4} \av{(A_1+A_1')(A_2-A_2')} _{\ket{\psi_{12}}}\av{A_3'}_{\ket{\psi_3}} \nonumber\\
  &   + \frac{1}{2}\av{(A_1-A_1')\otimes \openone_2}_{\ket{\psi_{12}}}|
  \end{align}
  Using the technique in (ii) above it is found that the maximum over $\ket{\psi_3}$ gives
\begin{align}
|\av{\mathcal{D}_3 ^{(i)}}|  \leq\, &
| \frac{\sqrt{2}}{4} \, \big( \,
 \av{(A_1+A_1')\,A_2} _{\ket{\psi_{12}}}^2\nonumber\\&
 +\av{(A_1+A_1')\,A_2'} _{\ket{\psi_{12}}}^2\,
 \big)^{1/2} \nonumber\\&+ \frac{1}{2}\av{(A_1-A_1')\otimes \openone_2}_{\ket{\psi_{12}}}  |.
\end{align}
Without losing generality we choose $A_i$, $A_i'$ in the $x-z$ plane \cite{toner} and  $\ket{\psi}_{12}=\cos\theta \ket{01}+\sin\theta\ket{10}$.  We can use the symmetry to set $A_1=A_2=A$ and $A_1'=A_2'=A'$. This gives
\begin{align}
|\av{\mathcal{D}_3 ^{(i)}}|  \leq &\,|\frac{1}{2}(a_z -a_z') \cos( 2\theta) +
\frac{\sqrt{2}}{4}\big( (a_z+a_z')^2+\nonumber\\
&((a_x+a_x')^2\sin(2\theta))^2 \big)^{1/2}|.
\end{align}
Since the observables $A$ and $A'$ must be orthogonal (i.e., $\vec{a}\cdot\vec{a}'=0$),
 this expression obtains its maximum for $a_x=a_x'=1/\sqrt{2}$ and $a_z=-a_z'=1/\sqrt{2}$. We finally get:
\beq
|\av{\mathcal{D}_3 ^{(i)}}|  \leq \frac{\sqrt{2}}{2}\cos( 2\theta) +\frac{1}{2}\sin( 2\theta)\leq \sqrt{3/4}.
\eeq

(iv)   We use the exact same steps of the proof of Sun \& Fei of \Eq{4}  (i.e., proof of
Theorem $2$ in \cite{sun}) but since the observables are orthogonal only case (4) of that proof needs
to be evaluated. This can be easily performed for the left hand side of \Eq{new2} that contains only two
terms instead of the three terms on the right hand side of \Eq{4}, thereby resulting in only a minor modification of the calculations \footnote{In further detail,
 the proof in Sun \& Fei \cite{sun} for the case of orthogonal observables amounts to  (using the terminology of their proof)  $|\vec{s}_i|=|\vec{t}_i|=\sqrt{2}/2$. Thus only step (4) needs to be evaluated and this gives
  $\omega=(\cos(\theta_1+\theta_2+\theta_3) -\sin(\theta_1+\theta_2+\theta_3))^2\leq2$.
As in the proof of Theorem 2 we have  $\omega=\av{\mathcal{D}_3^{(i)}} ^2+\av{\mathcal{D}_3^{(i+1)}}^2$ (i.e., the l.h.s. of \Eq{new2}),where again we have chosen $i =1$, but by symmetry the proof goes analogous for $i=2,3$. }  giving the result  $\av{D_{3}^{(i)}}^2 +\av{D_{3}^{(i+1)}}^2  \leq 2$. $\square$

These results for orthogonal observables can again be interpreted in terms of  the space given by the
coordinates $\mathcal{D}_3^{(i)}$ (for $i=1,2,3$). The same structure as in Fig. \ref{fig1} then arises
but with the different numerical bounds of Theorem 2.
The fully separable states are confined to a cube with edge length $\sqrt{3}$ and the biseparable
states in partition $j=1,2,3$ are confined to cuboids with size either$ \sqrt{6}\times\sqrt{3}\times\sqrt{3} ,~
\sqrt{3}\times \sqrt{6}\times\sqrt{3}$, or $\sqrt{3}\times\sqrt{3}\times\sqrt{6}$. Furthermore, all three-qubit
states are in the intersection of firstly the cube with edge length $\sqrt{6}$, secondly of  the three orthogonal
cylinders with radius $\sqrt{2}$, and thirdly of the sphere with radius $\sqrt{3}$.

\begin{figure}[!h]
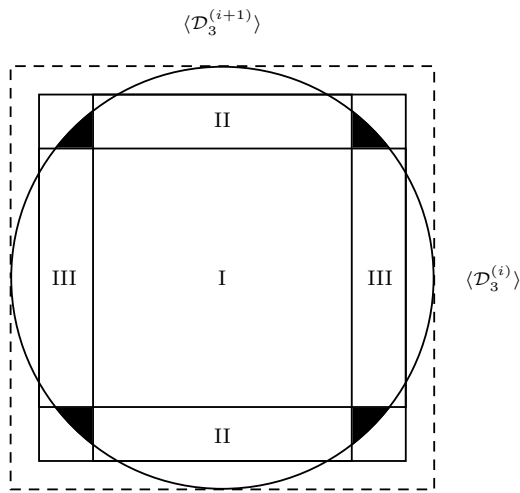

\pspicture(-3,-3)(3,4)
\psset{xunit=2,yunit=2}
\pswedge[fillstyle=solid,fillcolor=black](0,0){2.82}{38}{52}
\pswedge[fillstyle=solid,fillcolor=black](0,0){2.82}{128}{142}
\pswedge[fillstyle=solid,fillcolor=black](0,0){2.82}{218}{232}
\pswedge[fillstyle=solid,fillcolor=black](0,0){2.82}{308}{322}
\psframe*[linewidth=0.7pt,linecolor=white,fillcolor=white](-0.866,-1.225)(0.866,1.225)
\psframe*[linewidth=0.7pt,linecolor=white,fillcolor=white](-1.225,-0.866)(1.225,0.866)
\psframe[linewidth=0.7pt,fillcolor=white](-0.866,-1.225)(0.866,1.225)
\psframe[linewidth=0.7pt,fillcolor=white](-1.225,-0.866)(1.225,0.866)
\psframe[linewidth=0.7pt](-1.225,-1.225)(1.225,1.225)
\psframe[linestyle=dashed, dash=4pt 3pt,linewidth=0.7pt](-1.414,-1.414)(1.414,1.414)
\pscircle[linewidth=0.7pt](0,0){2.82}
\rput(0,0){\footnotesize{I}}
\rput(0,1.05){\footnotesize{II}}
\rput(0,-1.05){\footnotesize{II}}
\rput(1.05,0){\footnotesize{III}}
\rput(-1.05,0){\footnotesize{III}}
\rput(0.0,1.7){\scriptsize{$\av{\mathcal{D}_3^{(i+1)}}$}}
\rput(1.8,0){\scriptsize{$\av{\mathcal{D}_3^{(i)}}$}}
\endpspicture
\caption{The results of Theorem 3 for orthogonal observables. For comparison to the case
where the observables were not restricted to be orthogonal, the dashed square is included
that has edge length $2\sqrt{2}$ and which is the largest square in Fig. \ref{fig1}. }
\label{fig2}
\end{figure}

The corresponding $\mathcal{D}_3^{(i)}-\mathcal{D}_3^{(i+1)}$ plane  is drawn in Fig. \ref{fig2}.
Compared to the case where no restriction was made to orthogonal observables  (cf. Fig. \ref{fig1}) we see that we can
still distinghuish the different kinds of biseparable states, but they can still
not be distinghuished from
fully three-particle entangled states since both types of states still have the same maximum for
$\av{\mathcal{D}_{3}^{(i)}}$. Furthermore, the ratio of the different maxima of
$\av{\mathcal{D}_3^{(i)}}$ for fully separable  and bi-separable states is still the same,
i.e., the ratio is $\sqrt{2}/1=(\sqrt{3/2})/(\sqrt{3/4})=\sqrt{2}$.

The black areas in Fig \ref{fig2} are again non empty since states exist that have both $|\av{D_{3}^{(i)}}|>\sqrt{3/4}$ and $|\av{D_{3}^{(i+1)}}|>\sqrt{3/4}$ for the case of orthogonal observables. 
For example, the $W$-state of \Eq{W} gives $|\av{D_{3}^{(i)}}|=0.906$ for all $i$, for the local angles $\alpha_i= 0.54=\beta_i -\pi/2$ in the $x-z$ plane.

\section{Discussion}\noindent
Let us take another look at the quadratic inequalities $\av{D_{3}^{(i)}}^2 +\av{D_{3}^{(i+1)}}^2  \leq 5/2$ of
\Eq{new} for general observables and $\av{D_{3}^{(i)}}^2 +\av{D_{3}^{(i+1)}}^2  \leq 2$  of \Eq{new2}
for orthogonal observables. These can be interpreted as monogamy inequalities for maximal biseparable
three-particle quantum correlations (i.e., biseparable correlations that violate the inequalities maximally), since the inequalities show that a state that has maximal
bi-separable Bell correlations for a certain partition can not have it maximally for another partition.
Indeed, when
partition $i$ gives  $|\av{D_{3}^{(i)}}|= \sqrt{2}$ it must be the case according to  \Eq{new} that
for the other two partitions both $|\av{D_{3}^{(i+1)}}|\leq  \sqrt{1/2}$ and
 $|\av{D_{3}^{(i+2)}}|\leq \sqrt{1/2}$ must hold. The latter two must thus be non-maximal as soon as the first type of
 biseparable correlation is maximal. And for the second inequality of \Eq{new2} using orthogonal observables we get that when
 $|\av{D_{3}^{(i)}}|= \sqrt{3/2}$ (this is maximal) it must be the case that both
 $|\av{D_{3}^{(i+1)}}|\leq  \sqrt{1/2}$ and
 $|\av{D_{3}^{(i+2)}}|\leq \sqrt{1/2}$, which is non-maximal.

 From this we see that the first (i.e., \Eq{new} for general observables)
 is a stronger monogamy relationship than the second (i.e., \Eq{new2} for 
 orthogonal observables) since the trade-off
 between how much the maximal value for $|\av{D_{3}^{(i)}}|$ for one 
 partition $i$ restricts
 the value of $|\av{D_{3}^{(i+1)}}|$, $|\av{D_{3}^{(i+2)}}|$ 
 for the other two partitions below the maximal value  
 is larger in the first case than in the second case.
 
 Let us see how this compares to the monogamy inequality      
     $ \av{\mathcal{B}^{(i)}_2}^2 +\av{\mathcal{B}^{(i+1)}_2}^2\leq 2$  which was recently obtained by  Toner and Verstraete \cite{toner}. Note that $|\av{\mathcal{B}^{(i)}_2} |\leq 1$ is the ordinary Bell-CHSH inequality for the two qubits other than qubit $i$.  We see that this monogamy inequality is even more strong than the ones presented here, since  when  $|\av{\mathcal{B}^{(i)}_2}|$ obtains its maximal value of $\sqrt{2}$ it must be that
     $|\av{\mathcal{B}^{(i+1)}_2}|$=$|\av{\mathcal{B}^{(i+2)}_2}|=0$.

Furthermore, the monogamy relationship of Toner \& Verstraete shows that the so-called nonlocality that is indicated by correlations that violate the Bell-CHSH inequality \footnote{A nonlocal correlation is a correlation that can not be reproduced by shared randomness or any other local variables. It is detected by means of a violation of a Bell-type inequality.} cannot be shared (cf. \cite{scarani}): as soon as for some $i$ one has 
$|\av{\mathcal{B}^{(i)}_2}| > 1$, it must be that  both $|\av{\mathcal{B}^{(i+1 )}_2} |< 1$ and $|\av{\mathcal{B}^{(i+2 )}_2} |< 1$.  However, in Ref. \cite{collins} it was nevertheless shown that a bipartite Bell-type inequality exists
where it is the case that the nonlocality that this inequality allows for can be shared.
Since $|\av{D_{3}^{(i)}}|\leq1$ are Bell-type inequalities (i.e., local realism has to obey them, see \Eq{ineqN}) whose violation can be seen to indicate some nonlocality, the inequalities considered here could possibly also allow for some nonlocality sharing. 

Indeed, this is the case since it was shown that the black areas in Fig. \ref{fig1} are nonempty.
The Bell-type inequalities given here thus allow for sharing of the nonlocality of biseparable
three-particle quantum correlations that is indicated by a violation of these inequalities.
 
In conclusion, we have presented stronger bounds for bi-separable correlations in three-partite
systems than were given in \cite{sun} and extended this analysis to the case of the restriction to orthogonal observables
which gave even stronger bounds. The quadratic inequalities for biseparable correlations gave
a monogamy relationship for correlations that violate the inequalities maximally (i.e., these cannot be shared), but they nevertheless did allow for sharing of the non-maximally violating correlations.  

We hope that future research will reveil more of the structure of
the different kinds of partial separability in multipartite states and of the monogamy of multi-partite
Bell correlations.  It could therefore be fruitful to generalize this work from three to a larger number of parties.

\emph{Acknowledgements.}--- The author thanks Jos Uffink for fruitful discussions.

\end{document}